# Dirty Spin Ice: The Effect of Dilution on Spin Freezing in $Dy_2Ti_2O_7$

J. Snyder[1], J. S. Slusky[2], R. J. Cava[2], P. Schiffer[1*]

[1]*Department of Physics and Materials Research Institute, Pennsylvania State University, University Park PA 16802*

[2]*Department of Chemistry and Princeton Materials Institute, Princeton University, Princeton, NJ 08540*

## ABSTRACT

We have studied spin freezing in the diluted spin ice compound $Dy_{2-x}Y_xTi_2O_7$ where the non-magnetic Y ions replace the magnetic Dy ions on the frustrated pyrochlore lattice. Magnetic a.c. and d.c. susceptibility data are presented with an analysis of relaxation times for dilutions of $x$ = 0, 0.1, 0.2, and 0.4. Site dilution apparently decreases the relative number of spins participating in the ice-like freezing near 16 K while leaving the freezing temperature unchanged. Correspondingly the distribution of relaxation times associated with the freezing is broadened only slightly with increasing dilution, suggesting that the freezing process observed near T = 16 K involves the development of local correlations among the spins.

[*]*schiffer@phys.psu.edu*

Geometrically frustrated magnetic materials, in which the topology of the spin lattice leads to frustration of the spin-spin interactions, have recently been demonstrated to comprise a new class of magnets displaying unique cooperative spin states [1,2]. Although geometrical magnetic frustration has been most extensively studied in materials with antiferromagnetic nearest-neighbor interactions, the effects of strong frustration have also been found in the so-called "spin ice" materials (such as $Dy_2Ti_2O_7$, $Ho_2Ti_2O_7$, and $Ho_2Sn_2O_7$) [3,4,5,6,7,8,9,10,11,12] in which ferromagnetic and dipolar interactions can be frustrated [13,14,15,16]. The spins in these compounds are governed by the same statistical mechanics as the hydrogen atoms in the ground state of ordinary hexagonal ice (*Ih*) [17, 18,19]. In ice, oxygen ions reside at the center of tetrahedra with two of the four nearest hydrogen ions (protons) situated closer to it that the remaining two, as shown in figure 1a. In spin ice materials, the magnetic rare-earth ions are situated on a lattice of corner-sharing tetrahedra, and their spins are constrained by crystal field interactions to point either directly toward or directly away from the centers of the tetrahedra as shown in figure 1b. To minimize the dipole and ferromagnetic exchange interactions, the spins on each tetrahedron must be oriented such that two spins point inward and two point outward in exact analogy to the protons in ice.

The spin ice state has been demonstrated experimentally through neutron scattering studies [5,12,20] and also through measurements of the magnetic specific heat, which yield a measured ground state entropy in good agreement with the theoretical prediction for the "ice rules" (first codified by Pauling) and experimental results for ice [7,17,19]. While the spin entropy only freezes out below $T \sim 3$ K in $Dy_2Ti_2O_7$, magnetic susceptibility studies show a strongly frequency dependent cooperative spin-freezing at $T$

~ 16 K [8,9], and then a sharp drop at $T \sim 2$ K [21]. In contrast to traditional spin glasses, the T ~ 16K spin-freezing transition is associated with a very narrow range of relaxation times, and, rather than quenching the spin-freezing as in spin glasses, application of a magnetic field is found to enhance the spin ice freezing. The dynamic spin-ice freezing in $Dy_2Ti_2O_7$ is therefore a rather unusual example of glassiness in a magnetic system. Due to the purity of the system, it provides an excellent venue in which to study a simplified model of the complex thermodynamics of ice as well as the more general consequences of frustration in the limit of low disorder.

Here we report a study of this spin freezing in the diluted spin ice compound $Dy_{2-x}Y_xTi_2O_7$ where we introduce controlled disorder by substituting non-magnetic Y ions for the $J = 15/2$ $Dy^{3+}$ ions on the frustrated pyrochlore lattice. We find that such dilution decreases the relative number of spins participating in the ice-like freezing while leaving the freezing temperature, $T_f \sim 16$ K, and its frequency dependence unchanged. Correspondingly the distribution of relaxation times associated with the freezing is broadened only slightly with increasing dilution, suggesting that the freezing processes are essentially unchanged by such dilution. The combination of these observations strongly suggests that the spin freezing transition at T ~ 16K is cooperative in nature and involves the development of short-range spin-spin correlations.

Polycrystalline $Dy_{2-x}Y_xTi_2O_7$ samples were prepared using standard solid-state synthesis techniques described previously [8] with $x = 0, 0.1, 0.2,$ and $0.4$. The resulting powder was pressed into pellets from which pieces were cut to measure. X-ray diffraction demonstrated the samples to be of a single structural phase, and Curie-Weiss fits done to the high temperature susceptibility were consistent with $J = 15/2$ $Dy^{3+}$ ions.

To examine the development of the magnetic ground state of diluted spin ice, we study the magnetization ($M$) and the resultant d.c. susceptibility ($c_{dc} = dM/dH$) as well as the real and imaginary parts ($c'$ and $c''$) of the a.c. susceptibility ($c_{ac}$) as a function of temperature. The d.c. susceptibility measurements were performed with a Quantum Design MPMS SQUID magnetometer, and the a.c. susceptibility measurements were made with the ACMS option of the Quantum Design PPMS cryostat, calibrated with DyO.

Magnetic site dilution reduces the number of complete tetrahedra and hence changes the local magnetic landscape. As can be seen in Fig 1b, by removing a single spin from the complete Dy sublattice, two tetrahedra are broken and six Dy spins are left underconstrained. The extended effect of a single non-magnetic impurity suggests that a relatively low level of dilution would affect the nature of frustration, changing the number and types of relaxation processes available to the spins and also the freezing temperature. For this reason, dilution studies of antiferromagnetic geometrically frustrated magnets have been the focus of considerable recent attention [22,23,24]. In particular, $SrCr_{8-x}Ga_{4+x}O_{19}$ (SCGO), a frustrated antiferromagnet with a layered kagomé structure [25,26] has been studied extensively as a function of the concentration of the Cr ions relative to non-magnetic Ga. While not possessing the level of disorder traditionally associated with spin glasses, SCGO displays a spin glass transition at low temperatures ($T_f << Q_{Weiss} \sim 500$ K), which is associated with the geometrical frustration inherent to the kagomé lattice. Dilution would be expected to have a significant impact on such a strongly frustrated system, and indeed $T_f$ in SCGO is strongly reduced by dilution [27,28], decreasing by ~ 32% for $x = 2$ (~25% dilution).

In figure 2 we show the temperature dependent susceptibility of $Dy_{1.8}Y_{0.2}Ti_2O_7$ as a function of frequency to demonstrate the freezing phenomena typical to all the samples. We find that $c_{dc}$ increases monotonically with decreasing temperature as expected for a paramagnetic system with no spin freezing. While $c'(T)$ is virtually identical to $\chi_{dc}$ at our lowest frequency, as noted previously [8,9], at higher frequencies $c'(T)$ has a sharp decrease at $T_f \sim 16$ K, deviating well below $c_{dc}$. This sharp drop leads to a local maximum in $c'(T)$ correlated with a sharp rise in $c''(T)$ at a "freezing" temperature, $T_f$, which increases with frequency. This feature is a common signature of a glass transition in both structural glasses and spin glasses, and it is also observed in the dielectric permittivity of ice (which directly probes the local protonic motion) [29]. The drop in $c'(T)$ indicates that the spins' dynamic response is slowed such that they cannot respond to the time-varying magnetic field for $T < T_f$. This implies that the system is out of equilibrium on the time scale of the measurement, and thus that the observed properties depend on that time-scale – a classic signature of glassiness.

Figure 3 displays the real and imaginary parts of the a.c. susceptibility as a function of temperature for the different yttrium concentrations at a frequency of 1 kHz. These plots clearly show the suppression of freezing through dilution, since the drop in $c'$ at $T_f$ almost disappears between the pure sample and $x = 0.4$. Furthermore the rise in $c''$ at $T_f$ decreases over 40% from the pure to the $x = 0.4$ sample, indicating a reduction in absorption associated with the freezing process. Both of these changes are reflected at other frequencies as well, and they indicate that fewer spins are participating in the spin freezing at $T_f$ as the Dy sublattice is diluted. This implies that the T $\sim$ 16 K freezing is a collective phenomenon associated with the development of spin-spin correlations, rather than a single-ion effect which would not be affected by dilution.

Despite the strong reduction in the magnitude of the signature of freezing, the freezing process itself appears to be relatively unchanged in character by the dilution. The freezing temperature, $T_f$, as determined by the minimum in the derivative of $c''$, is unchanged to within ~2% across the span of dilutions. We can also fit the frequency dependence of the freezing temperature to an Arrhenius law, $f = f_0 e^{-E_a/k_B T_f}$ where $E_A$ is an activation energy for fluctuations and $f_o$ is a measure of the microscopic limiting frequency in the system. We plot such data in figure 4, and we find that $E_A$ is of order the single ion anisotropy energy and $f_o$ is of order MHz, both physically reasonable numbers for individual spin flips. Furthermore, the values of $E_A$ and $f_0$ (shown on the plot) do not vary systematically across the dilution series, and $E_A$ is actually constant to within a few percent. This suggests that magnetic site dilution has little impact on the mechanics of the spin freezing at T ~ 16 K, but simply acts to reduce the number of spins that participate in that transition.

Another method of parameterizing the spin freezing is through the Casimir-du Pré relation [30] which predicts, for a single relaxation time τ, that $c''(f) = f\tau\left(\dfrac{c_T - c_S}{1 + f^2 \tau^2}\right)$ where $c_T$ is the isothermal susceptibility in the limit of low frequency and $c_S$ is the adiabatic susceptibility in the limit of high frequency. We see in figure 5 that our data fit such a form at $T = 16$ K fairly well for all of the samples with a slight broadening with increasing $x$. The excellent fit for the $x = 0$ sample demonstrates the narrow range of relaxation times in the pure system [8]. Note that this behavior is in sharp contrast to that in other dense magnetic systems exhibiting glass-like behavior, in which the peak typically span several decades [31,32,33]. Notably, the peak is broadened only slightly in the diluted samples: the full-width at half maximum reaches only about 1.4 decades in frequency – still quite close to the theoretical expectation of 1.14 decades for a single relaxation mode. This slight broadening is further evidence, however, that the T ~ 16 K freezing is not a single-ion effect but rather that spin relaxation near that temperature is dependent on spin-spin interactions. The broadening of the peak can be also be parameterized by assuming a range of relaxation times, as has been demonstrated in dielectric materials [34], which is planned for a future study [35]. The quality of the fits can be examined further by a Cole-Cole (Argand) plot of $c'$ vs. $c''$ (figure 6). For the $x = 0$ sample, this is a close approximation of the theoretically expected semicircle corresponding to a single relaxation mode for the pure sample [8]. It does deviate more from semi-circular behavior with added dilution (although still not approaching the several decades of breadth seen in traditional spin glasses). The characteristic relaxation time, identified as the peak in $c''$ in this representation, can be seen to shift slightly to shorter times, from 0.3 ms in the pure sample to 0.2 ms in the $x = 0.4$ sample. This result indicates a somewhat faster relaxation processes in the diluted samples, which is consistent with a reduction in the local constraints on the spins. The rather small magnitude of the change suggests, however, that the relaxation processes are essentially unchanged with dilution.

The combined data yield two major results: dilution suppresses the magnitude of the freezing features in the susceptibility, but the quantitative parameters which characterize the spin freezing, e.g. $T_f$, $E_A$, and the characteristic timescales associated with the dynamics, are essentially unaffected by dilution even up to $x = 0.4$. One possible explanation for the suppression of spin-freezing in the diluted samples is that the chemical substitution is altering the lattice constant, and thereby changing the strength of the spin-spin interactions (either exchange or dipole). If this were the case, however, one would expect a change in both $T_f$ and the timescales associated with the freezing. Thus we conclude that the changes in spin freezing as a result of dilution must be in the number of spins which experience slowed dynamics below $T_f$. This suggests that the spin-spin correlations associated with the freezing at $T_f$ are short-ranged, i.e. the Dy spins which have a full set of 6 Dy nearest neighbors are still freezing but those which directly neighbor an Y impurity do not participate. Presumably those spins with second and further neighbor Y sites are only slightly affected (presumably through dipole interactions), explaining why the range of relaxation frequencies is somewhat broadened in the diluted samples and why the characteristic relaxation time becomes shorter (because all spins are somewhat less frustrated by the reduction in the dipole fields).

The data demonstrate that spin freezing in $Dy_2Ti_2O_7$ is quite different from that in traditional spin glasses and also quite different from the spin-glass-like transition observed in the geometrically frustrated antiferromagnet, SCGO, where $T_f$ is strongly affected by dilution. Since the spin-spin correlations in SCGO are also fairly short-range [36, 37], the distinction appears that in $Dy_2Ti_2O_7$, the freezing at $T_f$ represents the development of only nearest-neighbor correlations, and that longer range freezing occurs entirely at a lower temperature – presumably $T \sim 2K$ where the entropy is observed to freeze out.

The presence of two distinct features in the $c(T)$ is perhaps the most interesting aspect of the spin-ice freezing $Dy_2Ti_2O_7$, and is not seen in the other well established spin ice material, $Ho_2Ti_2O_7$ [21]. It has been suggested that only the lower temperature feature represents freezing into an ice-like spin state because it is closer in temperature to the freezing in the heat capacity and the fact that no anomaly is seen above 2 K in the d.c. susceptibility [9]. Since dilution affects the $T \sim 16$ K feature, we believe that feature cannot correspond to a single ion effect, but must be also associated with the development of spin-spin correlations and therefore is a precursor to the lower temperature "spin ice" freezing of the entropy. Furthermore, since the magnitude of the freezing signature but not the other parameters characterizing the magnetic relaxation at $T \sim 16$ K are strongly affected by dilution, it appears that the freezing at that temperature reflects only the development of short range correlations (since longer range correlations would be affected by dilution). We conclude that the "spin ice" freezing is most likely occurring in two stages: at $T \sim 16$ K, spins develop short-range correlations (presumably in local units of tetrahedra or larger) but they continue to fluctuate at low frequencies, and then long range freezing is achieved upon further cooling to below $T \sim 2$ K. This sort of two-stage freezing has also been seen in a.c. susceptibility studies of the spin-glass-like freezing of the geometrically frustrated garnet $Gd_3Ga_5O_{12}$ [38], and future investigations may show it to be a common feature in other geometrically frustrated magnetic materials.

## ACKNOWLEDGEMENTS

We gratefully acknowledge helpful discussions with D. A. Huse, M. Gingras and support from the Army Research Office PECASE grant DAAD19-01-1-0021.

**Figure Captions**

**Figure 1** Schematic representation of frustration in water ice and spin ice. **a**. In water ice each hydrogen is close to one or the other of its two oxygen neighbors, and each oxygen must have two hydrogen ions closer to it than to its neighboring oxygen atoms. **b**. With its ions restricted to point either directly toward or away from the centers of the tetrahedra, spin ice mimics the same frustration. The open circle represents a site dilution where a yttrium atom has replaced one of dysprosium.

**Figure 2** Temperature dependence of the real and imaginary parts of the magnetic susceptibility ($c'$ and $c''$) of $Dy_{1.8}Y_{0.2}Ti_2O_7$ in the absence of a d.c. magnetic field.

**Figure 3** Temperature dependence of the real and imaginary parts of the magnetic susceptibility ($c'$ and $c''$) of $Dy_{2-x}Y_xTi_2O_7$ at 1 kHz and in the absence of a d.c. magnetic field.

**Figure 4** Measurement frequency versus the inverse of the freezing temperature with fits to an Arrhenius law. The resulting fit parameters ($f_0$ and $E_A$) are also shown.

**Figure 5** The imaginary part of the magnetic susceptibility scaled to peak amplitude and frequency for $Dy_{2-x}Y_xTi_2O_7$ in the absence of a d.c. magnetic field. Also shown is the peak expected from the Casimir-du Pré relation assuming a single characteristic relaxation time for the system.

**Figure 6** Cole-Cole plot of the susceptibility data at several frequencies at $T = 16$ K. The semicircular character of the plot demonstrates the narrow distribution of relaxation times around a characteristic relaxation time. A flattening of the circle is evidence of a slight broadening in the relaxation time distribution.

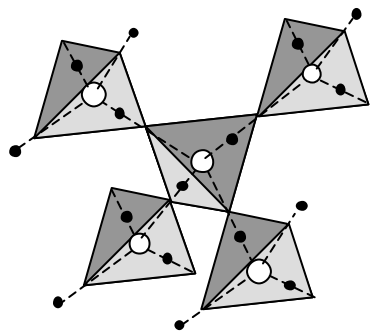 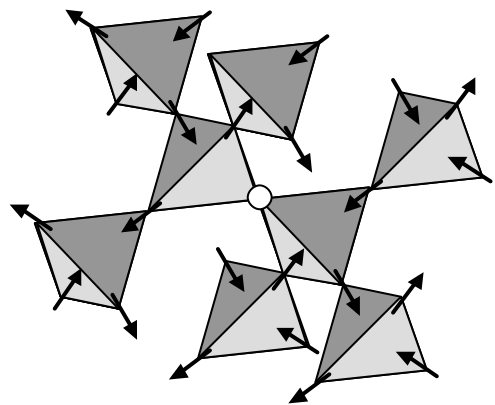

○ = $O^{2-}$
● = $H^+$

Ice                    Diluted Spin Ice

Figure 1

Snyder et al.

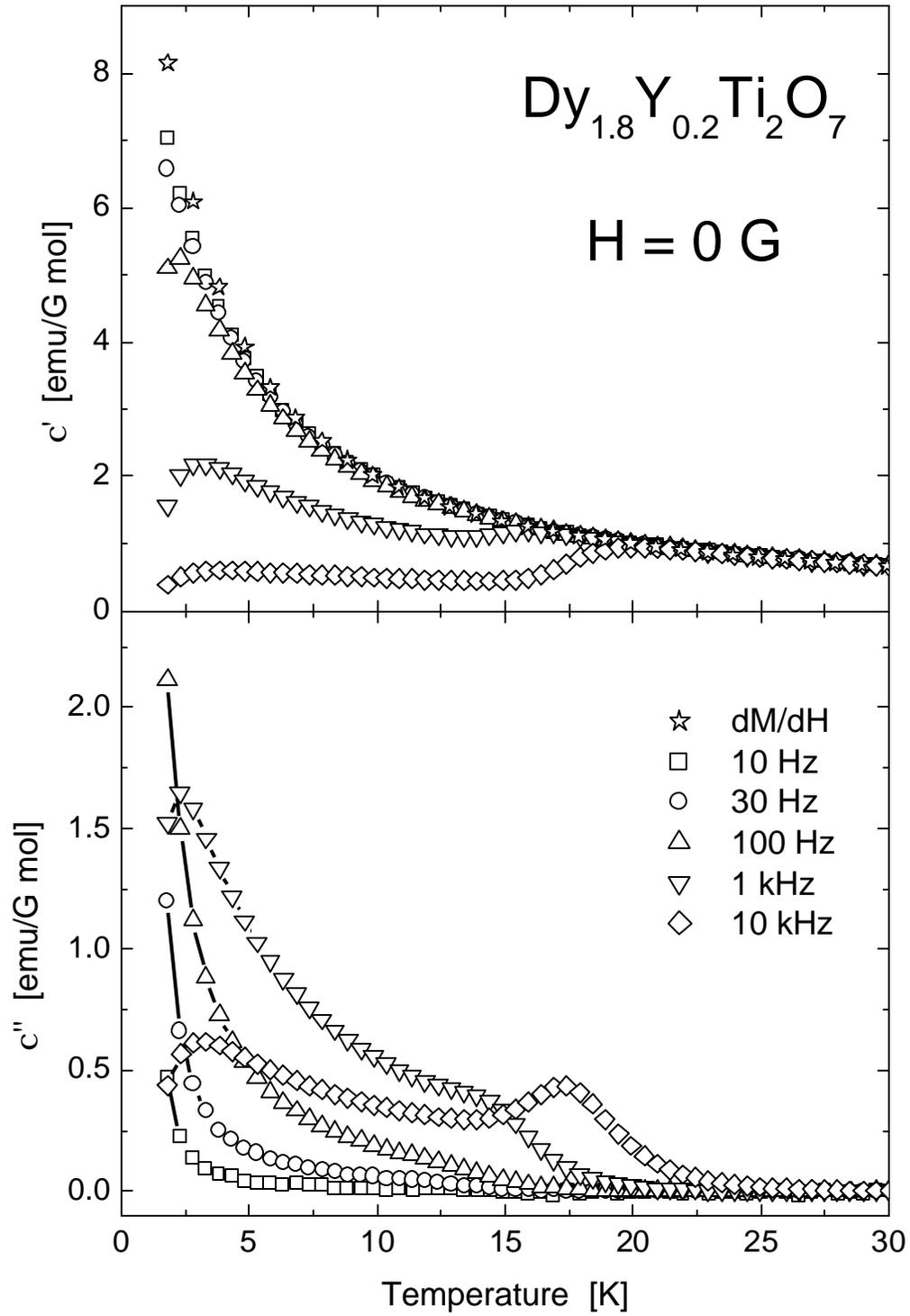

Figure 2

Snyder et al.

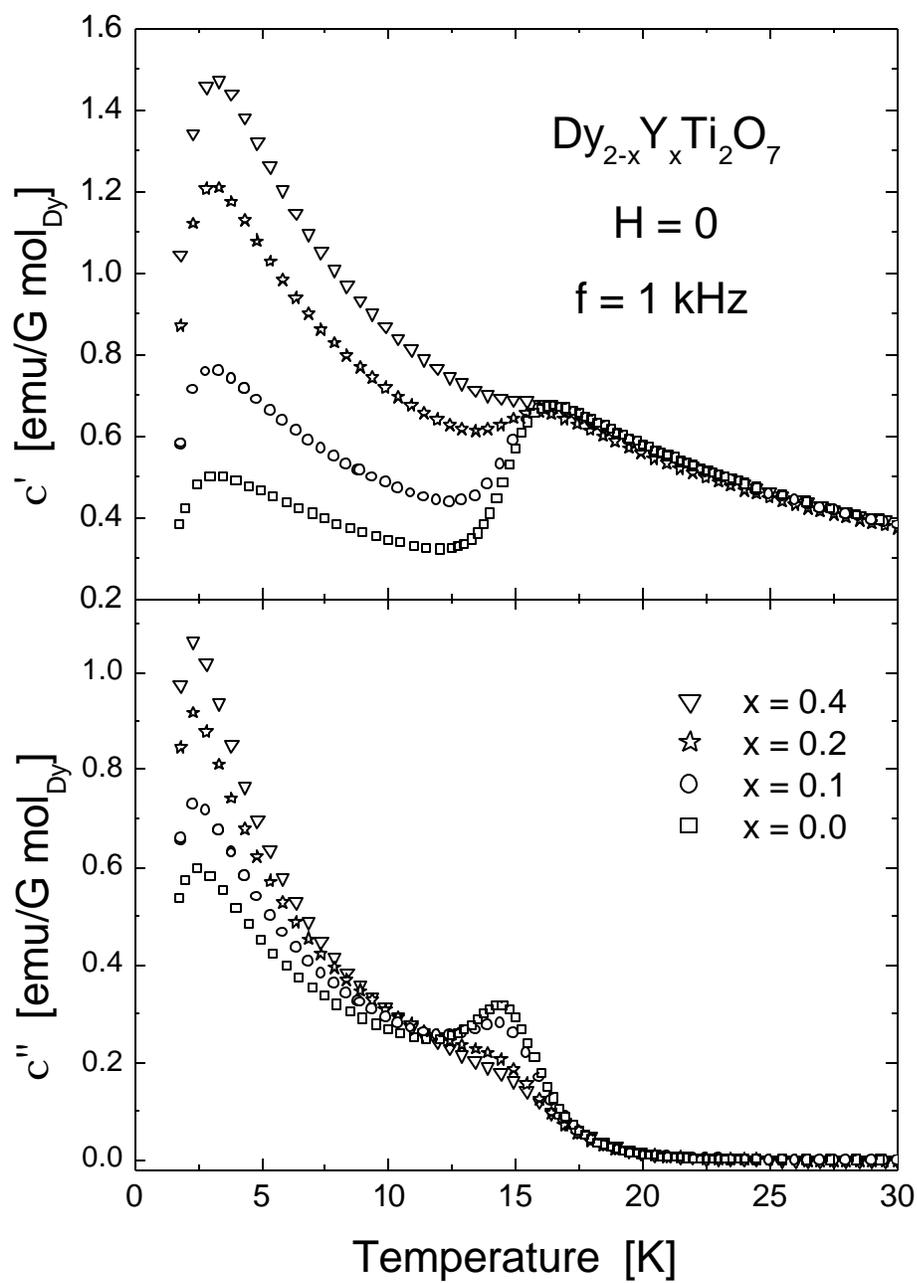

Figure 3

Snyder et al.

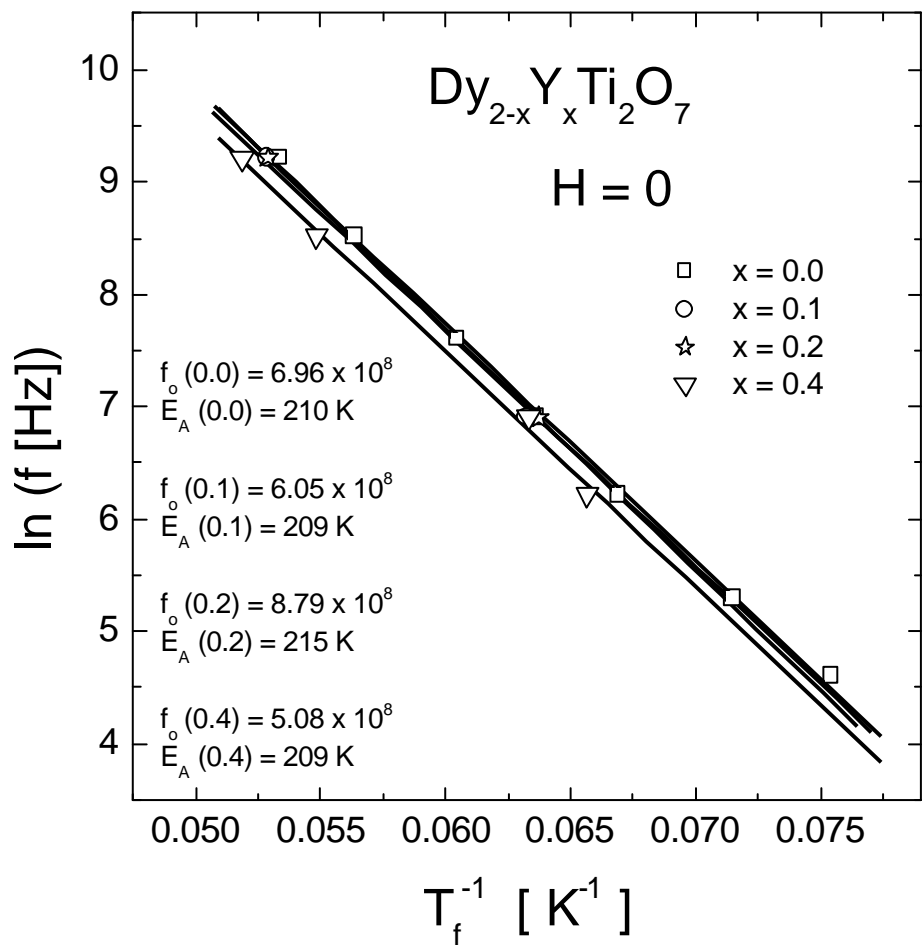

Figure 4
Snyder et al.

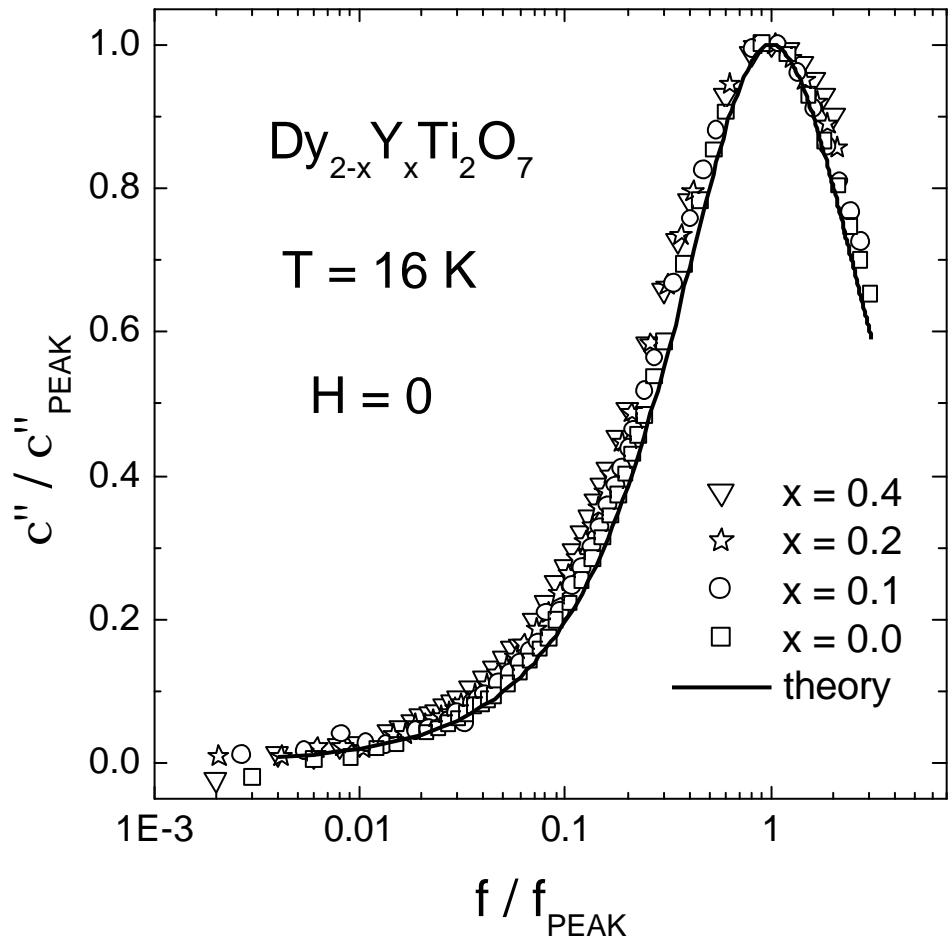

Figure 5

Snyder et al.

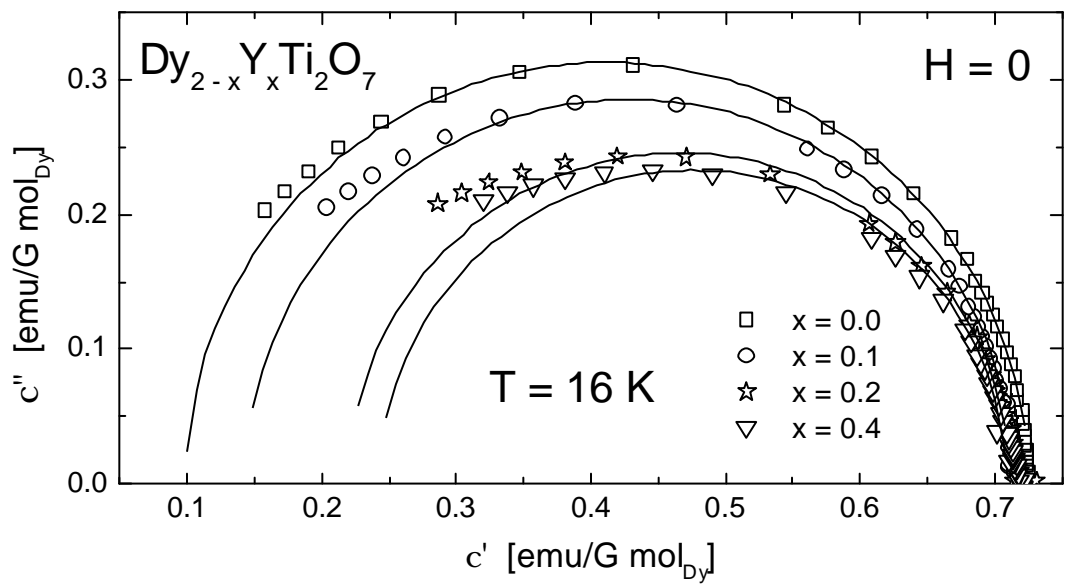

Figure 6

Snyder et al.

**References**

1. A.P. Ramirez, in *Handbook of Magnetic Materials* **13**, (ed. K. J. H. Buschow) (2001).

2. P. Schiffer and A.P. Ramirez, Comments Cond. Mat. Phys. **18**, 21 (1996).

3. S. T. Bramwell and M. J. P Gingras, Science **294**, 1495 (2001).

4. M.J. Harris, S.T. Bramwell, D.F. McMorrow, T. Zeiske, and K.W. Godfrey, Phys. Rev. Lett. **79,** 2554 (1997).

5. M.J. Harris, S.T. Bramwell, P.C.W. Holdsworth, and J.D.M. Champion, Phys. Rev. Lett. **81,** 4496 (1998).

6. S.T. Bramwell and M.J. Harris, J. Phys. Condens. Matt.. **10** L215 (1998).

7. A.P. Ramirez, A. Hayashi, R.J. Cava, R. Siddharthan, and B.S. Shastry, Nature **399,** 333(1999).

8. J. Snyder, J.S. Slusky, R.J. Cava, and P. Schiffer, Nature **413**, 48 (2001).

9. K. Matsuhira, Y. Hinatsu, and T. Sakakibara, J. Phys Condens. Matter **13**, L737 (2001).

10. K. Matsuhira, Y. Hinatsu, K. Tenya, and T. Sakakibara, J. Phys Condens. Matter **12,** L649 (2000) and J. Phys Condens. Matter **13**, L737 (2001).

11. A. L. Cornelius and J. S. Gardner, Phys. Rev. B **64** 060406 (2001).

12. H. Fukazawa, R. G. Melko, R. Higashinaka, Y. Maeno, and M. J. P. Gingras, Phys. Rev. B **65**, 054410 (2002).

13. R. Siddharthan, B. S. Shastry, A. P. Ramirez, A. Hayashi, R. J. Cava, and S. Rosenkranz, Phys. Rev. Lett. **83**, 854 (1999).

14. B.C. den Hertog and M.J.P. Gingras, Phys. Rev. Lett. **84**, 3430 (2000).

15. R.G. Melko, B.C. den Hertog, and M.J.P. Gingras, Phys. Rev. Lett. **87**, 067203 (2001).

16. R. Siddharthan, B. S. Shastry, and A. P. Ramirez, Phys. Rev. B **63**, 184412 (2001)

17. L. Pauling, *The Nature of the Chemical Bond* (Cornell Univ. Press, Ithaca, New York, 1945)

18. V.F. Petrenko and R.W. Whitworth, *Physics of Ice* (Clarendon, Oxford, 1999).

19. W.F. Giaque and J.W. Stout, J. Am. Chem. Soc. **58,** 1144 (1936).

20. S.T. Bramwell, M.J. Harris, B.C. den Hertog, M.J.P. Gingras, J.S. Gardner, D.F. McMorrow, A.R. Wildes, A.L. Cornelius, J.D.M. Champion, R.G. Melko, and T. Fennell, Phys. Rev. Lett. **87**, 047205 (2001).

21. For reasons which are unclear, this higher temperature freezing observed in $Dy_2Ti_2O_7$ [8, 9] is not seen in $Ho_2Ti_2O_7$. J. Snyder, J.S. Slusky, R.J. Cava, and P. Schiffer, unpublished.

22. P. Schiffer and I. Daruka, Phys Rev B **56**, 13712 (1997).

23. R. Moessner and A.J. Berlinsky, Phys. Rev. Lett. **83**, 3293 (1999)

24. A.J. Garcia-Adeva, D.L. Huber, Phys. Rev. B **64**, 014418 (2001).

25. A.P. Ramirez, G.P. Espinosa, and A.S. Cooper, Phys Rev Lett. **64,** 2070 (1990).

26. X. Obradors, A. Labarta, A. Isalgue, J. Tejada, J. Rodriguez, and M. Pernet, Solid State Commun. **65**, 189 (1988).

27. B. Martinez, F. Sandiumenge, and A. Rouco, Phys. Rev. B **46**, 10786 (1992).

28. P. Mendels, A. Keren, L. Limot, Phys. Rev. Lett. **85**, 3496 (2000), L. Limot et al., preprint (cond-mat/0111540).

29. S.J. Kawada, Phys. Soc. Jpn **44**, 1881 (1978).

30. H.B.G. Casimir and F.K. du Pré, Physica **5,** 507 (1938).

31. A.J. Dirkmaat, D. Huser, G.J. Nieuwenhuys, J.A. Mydosh, P. Kettler, and M. Steiner, Phys. Rev. B **36,** 352 (1987).

32. D. Huser, L.E. Wenger, A.J. van Duyneveldt, and J.A. Mydosh, Phys. Rev. B **27,** 3100 (1983).

33. C. Dekker, A.F.M. Arts, H.W. de Wijn, A.J. van Duyneveldt, and J.A. Mydosh, Phys. Rev. B **40,** 243 (1989).

34. See for example: R. J. Cava, R. M. Fleming, P. Littlewood, E.A. Rietman, L.F. Schneemeyer and R.G. Dunn, Phys. Rev. B **30**, 6 (1984) and R. J. Cava, R. M. Fleming, R. G. Dunn, E. A. Rietman, and L. F. Schneemeyer, Phys. Rev. B **32**, 4088 (1985).

35. J. Snyder, J.S. Slusky, R.J. Cava, and P. Schiffer, in preparation.

36. C. Broholm, G. Aeppli, and G. P. Espinosa, Phys. Rev. Lett. **65**, 3173 (1990).

37. C. Broholm, G. P. Espinosa, and A. S. Cooper, J. Appl. Phys. **67**, 5799 (1990).

38. P. Schiffer, A.P. Ramirez, D.A. Huse, P.L. Gammel, U. Yaron, D.J. Bishop, and A.J. Valentino, Physical Review Letters **74**, 2379 (1995).